\documentclass[aps, prl,pdf,superscriptaddress,amsmath,amssymb,amsfonts,twocolumn,showpacs,nofootinbib]{revtex4-2}

\usepackage{amssymb}
\usepackage{eqnarray,amsmath}
\newcommand{\abs}[1]{\left| #1 \right|} 
\usepackage{epsfig}
\usepackage{dcolumn}
\usepackage{bm}
\usepackage{braket}
\usepackage{balance}
\usepackage{physics}
\usepackage{csquotes}
\usepackage{mathtools}
\usepackage{graphicx,color,xcolor,colortbl}

\usepackage[colorlinks=true,
linkcolor=blue,
filecolor=blue,      
urlcolor=blue,
citecolor=blue]{hyperref}

\usepackage[normalem]{ulem}

\newcommand{\prlsection}[1]{%
  \noindent\textit{#1}---\ \ignorespaces}
\definecolor{Gray}{gray}{0.85}
\definecolor{LightCyan}{rgb}{0.88,1,1}

\newcolumntype{a}{>{\columncolor{Gray}}c}
\newcolumntype{b}{>{\columncolor{white}}c}

\begin{document}

\title{Self-Organized Stabilization of Straight Dark Solitons 
in Stripe Supersolids}

\author{Koushik Mukherjee}
\affiliation{Department of Engineering Science, University of Electro-Communications, Tokyo 182-8585, Japan}

\author{Hiroki Saito}
\affiliation{Department of Engineering Science, University of Electro-Communications, Tokyo 182-8585, Japan}
\begin{abstract}
Straight dark solitons in two-dimensional (2D) quantum fluids usually decay by transverse modulational instability, with no intrinsic suppression in contact-interacting Bose--Einstein condensates (BECs). We theoretically show that anisotropic long-range interactions in a quasi-2D dipolar BEC stabilize an embedded straight soliton, with spontaneous stripe order providing stronger pinning. The excitation spectra show that the lowest transverse solitonic branch remains gapped, while stripe-supersolid density modulation further hardens this branch and increases the soliton bending stiffness, penalizing transverse deformation. Accessible in current $^{166}$Er and $^{164}$Dy platforms, these results establish interaction-driven protection for straight dark solitons in structured quantum fluids.
\end{abstract}

%\begin{abstract}
%Straight dark solitons in two-dimensional (2D) quantum fluids are typically destroyed by transverse modulational instability, with no intrinsic suppression mechanism in contact-interacting condensates. Here we theoretically show that anisotropic long-range interactions in a quasi-2D dipolar Bose-Einstein condensate (dBEC) can stabilize an embedded straight dark soliton, while spontaneous stripe order provides an additional stronger pinning. Using extended Gross--Pitaevskii simulations and Bogoliubov--de~Gennes analysis, we find that dipolar anisotropy opens a frequency gap between anomalous and transverse solitonic branches, removing the mode collision that triggers decay; in the stripe-supersolid regime, crystalline density modulation further enhances this gap. From the excess energy associated with a stationary bent soliton, we demonstrate that the bending stiffness grows sharply with stripe contrast, penalizing transverse deformation. Accessible in current $^{166}$Er and $^{164}$Dy platforms, these results establish interaction-driven  protection for straight dark solitons in structured quantum fluids.
%\end{abstract}

%\date{\today}
\maketitle

\prlsection{Introduction}
Nonlinear excitations and topological defects govern transport, coherence loss, and relaxation across ordered media, from domain walls and flux lines in magnetic and superconducting materials to vortices and solitons in quantum fluids and nonlinear optical systems~\cite{Mermin1979,Klemen2008,Kivshar2003,Kevrekidis2015}.
A central question is whether a defect can be stabilized by the order it possesses, rather than by external pinning or confinement. Atomic Bose--Einstein condensates provide a clean setting for this question. In quasi-one-dimensional (1D) condensates, dark solitons, characterized by localized density depletions carrying a $\pi$ phase slip, have been generated and tracked experimentally~\cite{Burger1999,Denschlag2000,Becker2008,Fritsch2020}. In higher dimensions, however, they are generically unstable: transverse modulational instability~\cite{ZakharovRubenchik1973,Kuznetsov1988,Feder2000, Kivshar2000,Brand2002,Muryshev2002,Kevrekidisprl2017} drives decay into vortex rings in three dimensions (3D)~\cite{Anderson2001,Shomroni2009} and into solitonic vortices or vortex--antivortex structures in effectively two-dimensional (2D) geometries~\cite{Dutton2001,Donadello2014,Tamura2023}.

Previous stabilization strategies have relied on externally imposed mechanisms, including tight transverse confinement~\cite{Muryshev2002,Brand2002}, optical-lattice pinning~\cite{Kevrekidis2003,Theocharis2005}, and engineered spin-orbit coupling~\cite{Achilleos2013,Gallemi2016}. Long-range dipolar interactions offer a distinct route, in which the medium itself may generate the spatial order that protects the defect. Although dark solitons have been studied in 3D dipolar BECs~\cite{Nath2008}, stabilization there required an auxiliary external lattice rather than the dipolar fluid alone; stable dark solitons have also been reported in quasi-1D dipolar condensates~\cite{Bland2015,Edmonds2016}, where transverse decay is absent. Whether a genuinely 2D quantum fluid can self-organize a stabilizing landscape for a line phase defect remains open.

Ultracold strongly magnetic lanthanide gases provide such a setting~\cite{Lu2011DyBEC,Aikawa2012ErBEC, Ferrier-Barbut2016, Kadau2016, Schmitt2016, Chomaz2016}. Competition between contact and dipolar interactions drives a transition from a smooth  (SF) to a dipolar supersolid (SS), where phase coherence coexists with spontaneous crystalline density order~\cite{Gross1957, Yang1962, Andreev1969, Chester1970, Boninsegni2012}. Dipolar supersolidity, first realized in elongated geometries as 3D droplet arrays~\cite{Tanzi2019,Bottcher2019,Chomaz2019, biagioni2024measurement}, has since been explored mainly in trapped 3D condensates, including oblate geometries with transverse crystalline order~\cite{Norcia2021,Bland2022, Schmidt2021OblateRoton,Hertkorn2021Spectrum2D}. These systems have enabled studies of topological defects, primarily vortex structure and dynamics in crystalline backgrounds~\cite{Roccuzzo2020,Ancilotto2021,Klaus2022,Casotti2024,Mukherjee_Selective2025,SchubertPRR2025,Schubertprl2026}. In parallel, recent experimental advances in tightly confined quasi-2D dipolar gases~\cite{Wenzel2017,He2025BKT,Zhen2025ScaleInvariance} now enable access to stripe SS~\cite{Macia2012,Bombin2017,Cinti2019,Staudinger2023,Aleksandrova2024, Ripley2026,SanchezBaena2025, Poli2026Sound}, with possible experimental evidence~\cite{Wenzel2017,He2025StripeSupersolid} (see also Refs.~\cite{Li2017,Leonard2017, Chisholm2026Science}, for stripe states in spin-orbit coupled condensates). This raises a timely question: can anisotropic long-range interactions in a quasi-2D dipolar fluid stabilize a line phase defect, and can the resulting stripe order further reinforce this protection through a self-organized pinning landscape?

In this Letter, we show that it can. Using extended Gross--Pitaevskii equation (eGPE) simulations and Bogoliubov--de~Gennes (BdG) linear stability analysis, we show that anisotropic long-range interactions in a quasi-2D dipolar condensate can stabilize an embedded straight dark soliton against transverse decay. In our system, the dipoles are polarized in the $x$--$z$ plane at an angle $\alpha$ measured from the tightly confined $\hat{z}$ direction toward the soliton axis $\hat{x}$. Near in-plane polarization, this anisotropic dipolar field sharpens the soliton domain wall and hardens the lowest transverse solitonic mode, producing a stable window even before a finite stripe contrast develops. Upon increasing the relative dipolar interaction strength, spontaneous stripe order emerges and provides an additional crystalline pinning landscape, further widening the spectral gap and sharply enhancing the elastic stiffness coefficient $\sigma_2$ associated with soliton bending. This reveals a hitherto unexplored spectral and elastic mechanism for stabilizing straight dark solitons in a self-organized quantum fluid. The required parameters lie within current $^{166}$Er and $^{164}$Dy platforms, making interaction-driven stabilization of line defects experimentally accessible in structured quantum fluids~\cite{Chomaz2023, Recati2023, mukherjee2023droplets}.

\begin{figure}[t]
    \centering
    \includegraphics[width=\columnwidth]{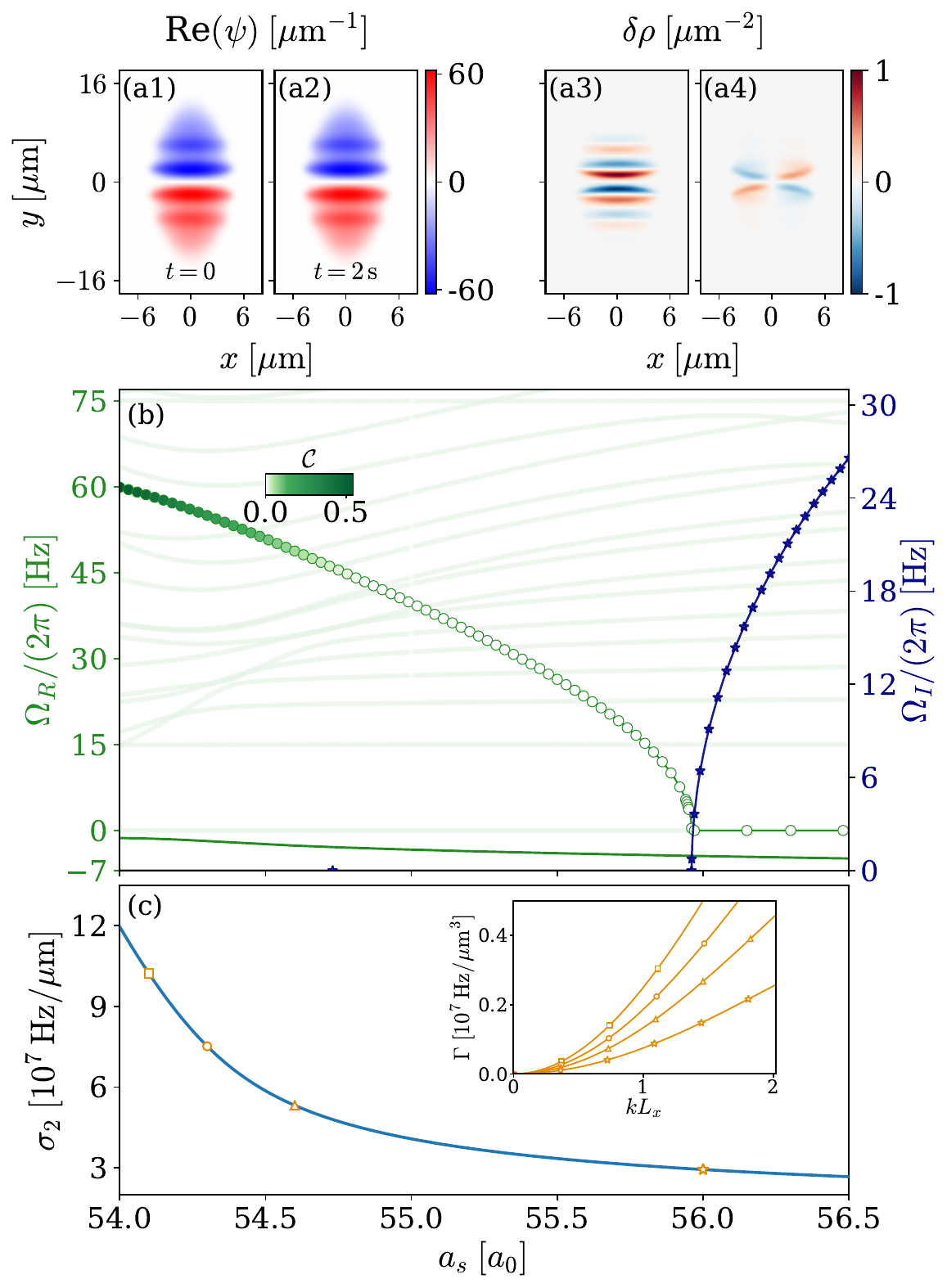}
  \caption{\textbf{Straight soliton stability
enhanced by SS order.}
(a1)~Real part of the SS wave function, $\mathrm{Re}(\psi)$, 
for the straight-soliton ground state and (a2)~at $t=2\,\mathrm{s}$ 
for $a_s = 54.20\,a_0$.
(a3)--(a4)~Stationary density fluctuations, 
$\delta \rho = 2\,\mathrm{Re}[\psi(\mathcal{U}+\mathcal{V}^*)]$, for the longitudinal and 
transverse collective modes, respectively.
(b)~Collective excitation spectrum versus $a_s$ for modes having positive Bogoliubov norm, $\mathcal{K} \ge 0$. 
Green branches denote real mode frequencies $\Omega_{\rm R}/(2\pi)$; 
the blue branch denotes the imaginary component $\Omega_{\rm I}/(2\pi)$, 
marking the onset of dynamical instability. Marker fill encodes the density contrast $\mathcal{C}$ of the ensuing SS state.
Pale points show the full excitation spectrum.
(c)~Elastic stiffness coefficient $\sigma_2$ versus $a_s$.
Inset: stiffness dispersion $\Gamma(k)$ versus bending 
wavevector $k$ for selected $a_s$ values corresponding to the marked 
points in~(c). 
All results are for $N = 40\,000$ atoms at polarization angle 
$\alpha = \pi/2$.}
    \label{fig:fig1_manuscript}
\end{figure}
\prlsection{Model}
We model a quasi-2D dipolar Bose gas under tight harmonic confinement along $\hat{z}$, with condensate wave function $\psi(x,y)$ in the transverse plane. Its dynamics obeys an eGPE containing contact interactions, nonlocal dipole--dipole interactions, and the Lee--Huang--Yang (LHY) correction~\cite{Lee1957, Schutzhold2006, Lima2011, *Lima2012, *Wachtler2016} in the local-density approximation. The contact interaction coefficient $g$ is set by the tunable $s$-wave scattering length $a_s$~\cite{chin2010}, while the dipolar coefficient $g_{\rm dd}$ is fixed by the dipolar length $a_{\rm dd}$. The LHY coefficient $g_{\rm LHY}$ depends on the relative dipolar strength $\epsilon_{\rm dd}=a_{\rm dd}/a_s$.
%The dimensionless contact and dipolar interaction strengths are
%$g=\sqrt{8\pi\lambda}\,Na_s/l_0$ and
%$g_{\rm dd}=\sqrt{8\pi\lambda}\,Na_{\rm dd}/l_0$,
%where $l_0=\sqrt{\hbar/M\omega_z}$ and $\lambda=\omega_z/\omega_y$.
The nonlocal dipolar potential is evaluated in momentum space as
$\tilde{\Phi}_{\rm dd}(\mathbf{k})=\tilde{U}^{\rm 2D}_{\rm dd}(\mathbf{k})\tilde{n}(\mathbf{k})$,
with $\tilde{U}^{\rm 2D}_{\rm dd}$ retaining the full $\alpha$-dependent anisotropy~\cite{Fischer2006, Ronen2006}.

Stationary straight-soliton states $\psi_0$ are obtained by imaginary-time propagation of the eGPE while enforcing
$\psi(x,y)=\mathrm{sgn}(-y)|\psi(x,y)|$
after each iteration, which pins a nodal line at $y=0$ and imposes a $\pi$ phase jump across the soliton axis [see Fig.~\ref{fig:fig1_manuscript}(a1)]~\cite{Frantzeskakis2010,Kevrekidis2015}. The excitation spectrum is obtained by linearizing around $\psi_0(\mathbf{r})e^{-i\mu t}$ with
$\delta\psi\,e^{i \mu t}=\mathcal{U}(\mathbf{r})e^{-i\Omega t}+\mathcal{V}^{*}(\mathbf{r})e^{i\Omega^{*} t}$,
yielding BdG modes $(\mathcal{U},\mathcal{V})$ and complex frequencies $\Omega=\Omega_R+i\Omega_I$, with associated density fluctuation
$\delta \rho=2\mathrm{Re}[\psi_0(\mathcal{U}+\mathcal{V}^*)]$. The $\mu$ denotes chemical potential of the system.
Modes with $\Omega_I>0$ are dynamically unstable; if the spectrum contains a mode with negative real frequency, $\Omega_R<0$, and nonnegative BdG norm, $\mathcal{K}=\int d\mathbf{r}\,(|\mathcal{U}|^2-|\mathcal{V}|^2)> 0$, the state is energetically unstable~\cite{Svidzinsky2000,Muryshev2002, Kevrekidis2015}.

Throughout this work we use recent experimentally relevant $^{166}$Er parameters~\cite{Zhen2025ScaleInvariance, He2025StripeSupersolid} with $a_{\rm dd}=65.5\,a_0$ and,
$(\omega_x,\omega_y,\omega_z)/2\pi=(75,15,1100)\,\mathrm{Hz}$.
We choose the trap to be elongated along \(\hat{y}\), so that, for smaller $a_s$ and near in-plane polarization along $\hat{x}$ $(\alpha\approx\pi/2)$, the density modulation develops along $\hat{y}$ and forms stripes parallel to the soliton. We vary $a_s$ and $\alpha$ to assess the roles of anisotropic interactions and stripe-SS order in soliton stability.

\prlsection{Spectral signatures of stable straight dark solitons}
We first consider the maximally anisotropic case, $\alpha=\pi/2$, with $N=4.0\times10^4$ atoms. Figure~\ref{fig:fig1_manuscript}(b) shows the BdG excitation spectrum as a function of scattering length $a_s$, including both real and imaginary frequency components, $\Omega_{\rm R}$ and $\Omega_{\rm I}$, respectively, for $\mathcal{K} \ge 0$. The full spectrum contains the zero-frequency Goldstone mode associated with broken $U(1)$ symmetry, the center-of-mass dipole modes fixed by the trap frequencies, and higher collective excitations shown as pale background points. In particular, we focus on two solitonic excitation branches. The first is a soliton-localized branch (dark green line without markers in Fig.~\ref{fig:fig1_manuscript}(b)]) whose density fluctuation $\delta\rho$ is odd under $y\to -y$ and has a nodal line along the soliton core; see Fig.~\ref{fig:fig1_manuscript}(a3). This mode corresponds to a rigid displacement of the soliton core and is therefore the translational zero frequency mode in a homogeneous system; the trap shifts it to finite negative frequency.

The second, and most relevant for stability in our system, is the lowest transverse excitation. Its density fluctuation $\delta\rho$ is odd under $x\to -x$ and has a single nodal line perpendicular to the soliton axis [Fig.~\ref{fig:fig1_manuscript}(a4)]. This transverse mode drives  the decay of a straight soliton.
 As $a_s$ is increased, this  branch crosses other excitations, while retaining  its distinct density-fluctuation characters. Then, it softens completely at $a_s \simeq 55.967$, after which a finite imaginary component $\Omega_{\rm I}>0$ appears, signaling dynamical instability of the straight soliton. Lowering $a_s$ increases the relative strength of dipolar interactions, which sharpens the soliton domain wall and keeps the transverse branch at finite real frequency even before a finite stripe contrast develops.
With further decrease of $a_s$, a stripe density modulation emerges and reinforces this transverse-mode stabilization. We quantify the stripe order by the density contrast
$
\mathcal{C}=(n_{\rm max}-n_{\rm min})/(n_{\rm max}+n_{\rm min}),
$
computed from the
$n_{0}(y)=\int dx\,n_0(x,y)$
along the modulation direction. Here $n_{\rm max}$ is the second density maximum away from the soliton core, and $n_{\rm min}$ is the intervening minimum between adjacent maxima. In Fig.~\ref{fig:fig1_manuscript}(b), the $\Omega_{\rm R}$ markers are color-coded by $\mathcal{C}$, distinguishing the unmodulated SF regime $(\mathcal{C}=0)$ from the stripe-SS regime $(\mathcal{C}>0)$.
The emergent stripe modulation acts as a self-generated pinning landscape for the straight dark soliton. It strongly stabilizes the soliton by further hardening the lowest transverse solitonic branch and suppressing its softening toward zero frequency. This mechanism persists across different numbers of stripe sites, while remaining absent in contact-interacting BECs; see Appendices A and B in the \textit{End Matter}. Demonstrating this intrinsic stabilization mechanism for straight solitons in dipolar BECs is the central result of this work. Furthermore, the spectral stabilization is confirmed by real-time propagation of the stationary state for $t=2\,\mathrm{s}$ [Figs.~\ref{fig:fig1_manuscript}(a1)--(a2)], during which the straight soliton remains structurally intact.

\prlsection{Energetic cost of soliton bending} The BdG formalism based spectral analysis above establishes a dynamical criterion for soliton 
stability. To further understand this within an analytical model, we now ask the complementary 
question: what is the excess energy cost of a stationary transverse deformation at wavenumber $k$, and how does this cost increase with SS density modulation? 
In the trap center, where the background varies slowly along $x$, we approximate the soliton as a straight line of length $L_x$, $x \in [-L_x/2,\, L_x/2]$, and consider only the transverse harmonic confinement $V(y)\propto y^2$. Then, we parametrize a weak transverse bending of the soliton by a displacement field $u(x)$~\cite{Kevrekidisprl2017, Kevrekidis2018}. For $|u(x)|\ll \ell_T$ and $|\partial_xu|\ll1$, where $\ell_T$ denotes the
shortest transverse length scale over which $n_0$ varies, the deformed density is
  $n(x,y)=n_0\!\left(x,y-u(x)\right)
    = n_0+\delta n(x,y)+\mathcal{O}(u^3)$,
where 
%\begin{equation}
$  \delta n(x,y) \simeq - u(x)\,\partial_y n_0 + (u^2(x)/2)\,\partial_y^2 n_0$.
%        \label{eq:bent_soliton_density}
%\end{equation}
 Taking $u(x)=A\sin(kx)$, the resulting energy difference, $\Delta E=E(n_0+\delta n)-E(n_0)$ can be  calculated (details in the Supplemental Material) as  $\Delta E = A^2 L_x \Gamma(k)$, where the $\Gamma (k)$ is given by
%
%\begin{widetext}
%\begin{equation}
%\begin{aligned}
%\Gamma(k)
%&=
%\left(1-\frac{\sin(kL_x)}{kL_x}\right)
%\Bigg[
%\frac{\mathcal{I}_1}{4}
%+\frac{g\,\mathcal{I}_3}{4}
%+\frac{3\,g_{\rm LHY}\,\mathcal{I}_2}{8}
%\Bigg]
%+\frac{k^2}{4}\left(1+\frac{\sin(kL_x)}{kL_x}\right)\mathcal{I}_4
%+ \frac{g_{\rm dd}}{(2A^2L_x)}
%\int\frac{d\mathbf{k}}{(2\pi)^2}\,
%\widetilde{U}_{\rm dd}^{\rm 2D}(\mathbf{k})\,\widetilde{u}^2(k_x)k^2_y
%\widetilde{n}^2_0(k_y)
%\end{aligned}
%\label{eq:Gamma}
%\end{equation}
%\end{widetext}
%
\begin{widetext}
\begin{equation}
\begin{aligned}
\Gamma(k)
 =
& \frac{\hbar^2k^2}{4M}\left[1+\frac{\sin(kL_x)}{kL_x}\right]\int dy\,(\partial_y\sqrt{n_0})^2
 -
\left[
1-\frac{\sin(kL_x)}{kL_x}
\right]
\bigg\{
\frac{2\hbar^2}{M} \int dy\,
\left[
\left(\partial_y n_0^{1/4}\right)^4
+
n_0^{1/4}
\left(\partial_y n_0^{1/4}\right)^2
\partial_y^2 n_0^{1/4}
\right] \\  
& - \frac{1}{4} \int dy\, V(y)\,\partial_y^2 n_0(y) \bigg \}+ \frac{g_{\rm dd}}{(2A^2L_x)}
\int\frac{dk_x dk_y}{(2\pi)^2}\,
\widetilde{U}_{\rm dd}^{\rm 2D}(k_x, k_y) \bigg\{ \mathrm{Re}\!\left[\widetilde{n}_0^*(k_y)\,\widetilde{\delta n}(k_x,k_y)\right] + |\widetilde{\delta n}(k_x,k_y)|^2 \bigg\}.
\end{aligned}
\label{eq:Gamma}
\end{equation}
\end{widetext}
%
%with $\mathcal{I}_1 = \int dy\,(\partial_y^2\sqrt{n_0})^2$,
%$\mathcal{I}_2 = \int dy\,\sqrt{n_0}\,(\partial_y n_0)^2$,
%$\mathcal{I}_3 = \int dy\,(\partial_y n_0)^2$,
%$\mathcal{I}_4 = \int dy\,(\partial_y\sqrt{n_0})^2$.
with \enquote{tilde} denoting the Fourier transform of the quantity.
The first two terms involve weighted integrals of the transverse gradients 
of $n_0$ and are therefore directly sensitive to density modulations 
perpendicular to the soliton; the final term is the nonlocal 
dipolar contribution to the bending stiffness.
Following the standard long-wavelength elastic expansion for line and interface deformations~\cite{Safran1994}, we write
\begin{equation}
\Gamma(k)=\sigma_2 k^2+\sigma_4 k^4+\mathcal{O}(k^6),
\end{equation}
where $\sigma_2$ is the effective bending stiffness of the soliton line.

Figure~\ref{fig:fig1_manuscript}(c) shows the stiffness
$\sigma_2$ as a function of $a_s$, together with the dispersions
$\Gamma(k)$ [inset of Fig.~\ref{fig:fig1_manuscript}(c)], at representative values of $a_s$; see the markers. Comparison with Fig.~\ref{fig:fig1_manuscript}(b)
shows that, in the  SF regime $(\mathcal{C}=0)$, $\sigma_2$
is small and only weakly dependent on $a_s$. In contrast, as  $a_s$ decreases, the emerging periodic modulation produces 
a pronounced enhancement of $\sigma_2$. Physically, this stiffening 
reflects the intrinsic periodic landscape of the stripe SS, 
which penalizes lateral soliton displacements and thereby imposes a 
large static energy cost on bending deformations, complementing the spectral stabilization of straight solitons 
deep in the SS phase.

\begin{figure}[t]
    \includegraphics[width=0.99\columnwidth]{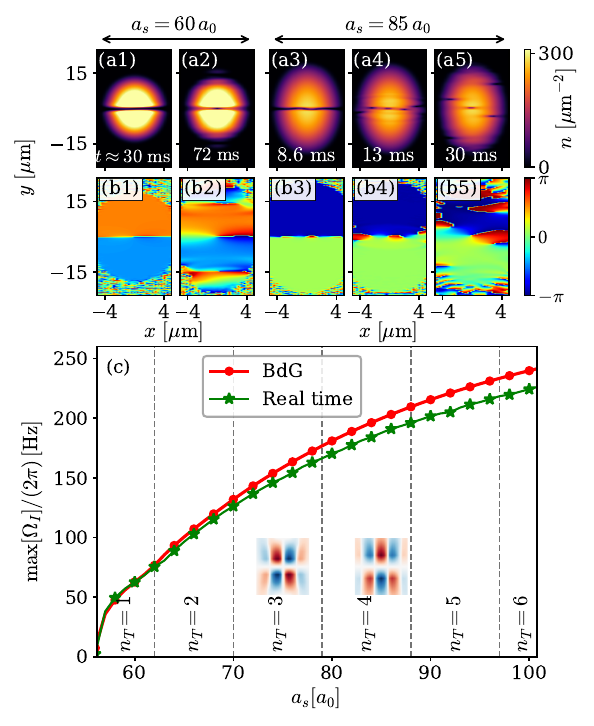}
\caption{
\textbf{Dynamical signatures of transverse instability}.
(a1)--(a2) Snapshots of the density profile at selected times (see legends) for scattering length $a_s=60\,a_0$, showing the nucleation of a solitonic vortex.
(a3)--(a5) Density profiles for $a_s=85\,a_0$, showing the bending of the straight soliton and the nucleation of vortex--antivortex pairs.
(b1)--(b5) Corresponding phase profiles, showing the phase winding associated with the density deformation.
(c) Maximum imaginary frequency, or instability growth rate, as a function of $a_s$ from BdG theory (red circles) and real-time dynamics (green stars).
Dashed vertical lines mark changes in the number $n_T$ of transverse nodal lines of the most unstable mode; insets show representative $\delta \rho$ patterns. Other parameters are as in the main text.
}  \label{fig:fig2_manuscript}
\end{figure}

\prlsection{Unstable soliton and vortex nucleation}
To identify nonlinear structures emerging from the unstable straight
soliton in the regime $a_s \gtrsim 55.96\,a_{0}$, we perform real-time eGPE simulations starting from the
stationary state perturbed as $\psi(\mathbf{r}) \to
\psi(\mathbf{r})[1 + \epsilon(\mathbf{r})]$, where $\epsilon(\mathbf{r})$
is a white-noise field with $|\epsilon| \le 10^{-4}$.
The dynamical outcome depends sensitively on proximity to the instability
threshold, yielding two distinct decay scenarios illustrated in
Figs.~\ref{fig:fig2_manuscript}(a1)--(a5) and (b1)--(b5).

Near threshold, only one  BdG mode has \(\Omega_{\rm I}>0\),
corresponding to an unstable transverse mode with one nodal line  ($n_T=1$) along
 $\hat{y}$ [Fig.~\ref{fig:fig2_manuscript}(c)].
This mode is selectively amplified during the dynamics,
generating a localized $2\pi$ phase slip at the center and nucleating
a single solitonic vortex (SV), as confirmed by the density and phase
profiles at $a_s = 60\,a_0$
[Figs.~\ref{fig:fig2_manuscript}(a2) and (b2)]. Here, unlike in Ref.~\cite{Brand2002}, the soliton-to-SV transition is interaction-controlled rather than confinement-driven.
For $a_s \gtrsim 62\,a_0$, however, we observe a qualitatively different dynamical behavior, as several transverse BdG modes ($n_T > 1$)
are simultaneously unstable.
The resulting higher-transverse-mode bending creates alternating high- and low-density regions along the straight soliton [Fig.~\ref{fig:fig2_manuscript}(a3)] and locations of $2\pi$ phase windings [Figs.~\ref{fig:fig2_manuscript}(b3)--(b6)]. This process nucleates multiple vortex--antivortex pairs, which propagate into the bulk during the dynamics [Figs.~\ref{fig:fig2_manuscript}(a3)--(a4)], a hallmark signature of the snake instability~\cite{ZakharovRubenchik1973}.

To establish a quantitative link between the BdG analysis and
the real-time dynamics, we extract the dominant growth rate
$\gamma_{\rm rt} = \max[\Omega_{\rm I}]$ directly from real-time
simulations via exponential fitting of the peak of the dynamical structure factor
$S(\mathbf{k}, t)$ computed from time-evolving density snapshots.
As shown in Fig.~\ref{fig:fig2_manuscript}(c), the extracted growth
rates are in excellent quantitative agreement with the BdG imaginary
frequencies $\Omega_{\rm I}$, directly identifying the leading unstable
collective mode that governs the early-stage dynamics.

\begin{figure}[t]
    \centering
    \includegraphics[width=\columnwidth]{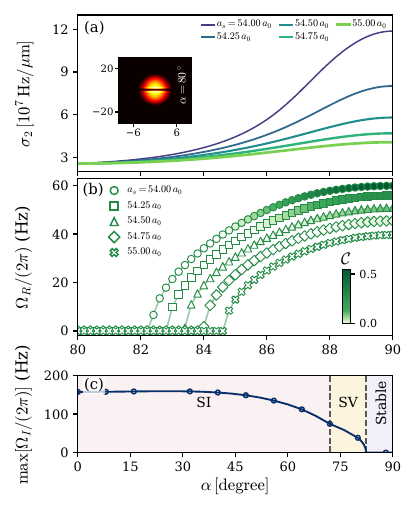}
\caption{\textbf{Tuning the polarization angle $\alpha$:} 
(a) Elastic stiffness coefficient $\sigma_2$ as a function of $\alpha$ for
different scattering lengths $a_s$. Inset: stationary density profile of the straight soliton for $\alpha=80^\circ$.
(b) Real frequency $\Omega_{\rm R}$ of the transverse soliton mode as a
function of $\alpha$ for different $a_s$. Markers are color-coded by the
density contrast $\mathcal{C}$. 
(c) Imaginary frequency $\Omega_{\rm I}$ as a function of $\alpha$. The stable,
solitonic-vortex (SV), and snake-instability (SI) regions are indicated. The SV and SI instability regions stem from the $n_T = 1$ and $n_T >1$ transverse modes, respectively. Other
parameters are as in the main text.
}   \label{fig:fig3_manuscript}
\end{figure}

\prlsection{Role of dipole orientation in soliton stabilization}
The polarization angle $\alpha$ controls the interaction anisotropy and therefore provides a direct tuning knob for soliton stability. At full in-plane polarization, $\alpha=\pi/2$, the dipoles are oriented along the straight-soliton axis $\hat{x}$, producing attractive head-to-tail interactions along $\hat{x}$ and predominantly repulsive interactions along the transverse direction $\hat{y}$. This anisotropy sharpens the soliton domain wall and localizes density near it, already contributing to the stable window below the threshold at $a_s \approx 55.96\,a_{0}$. It also drives density modulation along $\hat{y}$, trapping the soliton in a self-induced channel parallel to the SS stripes and, through Eq.~\eqref{eq:Gamma}, amplifying transverse density gradients and hence the bending stiffness $\sigma_2$. Tilting the dipoles away from $\alpha = \pi/2$ weakens the in-plane anisotropy,
suppresses the stripe modulation, and eventually restores a smooth SF
background; see the inset of Fig.~\ref{fig:fig3_manuscript}(a). The resulting flattening of $n_0(y)$ reduces the energetic penalty for bending and leaves the soliton susceptible to the transverse instability.

Figure~\ref{fig:fig3_manuscript} confirms this picture. The stiffness $\sigma_2$ is maximal at $\alpha=\pi/2$ and falls sharply as $\alpha$ decreases, becoming strongly suppressed for $\alpha < 82^{\circ}$ with negligible $a_s$ dependence [Fig.~\ref{fig:fig3_manuscript}(a)]. Correspondingly, the transverse-mode frequency $\Omega_{\rm R}$ follows the loss of density contrast $\mathcal{C}$ and softens rapidly [Fig.~\ref{fig:fig3_manuscript}(b)], eventually acquiring a finite imaginary component $\Omega_{\rm I}>0$ [Fig.~\ref{fig:fig3_manuscript}(c)]. The unstable region again separates into two regimes (as in Fig.~\ref{fig:fig2_manuscript}): a near-threshold sector with single solitonic-vortex (SV) nucleation and a sector producing multiple vortex--antivortex pairs. Robust stability persists only for $\alpha\gtrsim80^\circ$, with maximum stabilization at $\alpha=\pi/2$.

\prlsection{Conclusions and Outlook}
In summary, we have shown that anisotropic long-range interactions in a quasi-2D dipolar condensate alone can stabilize a straight dark soliton against transverse decay. Near in-plane polarization sharpens the soliton domain wall and prevents the lowest transverse branch from softening into an imaginary-frequency mode. In the stripe-SS regime, spontaneous density modulation forms a self-generated pinning landscape that enhances transverse density gradients, increases the bending stiffness $\sigma_2$, and confines the soliton along the stripe channel, an effect absent in purely contact-interacting systems. More broadly, this establishes a self-organized route to defect stabilization: unlike vortex pinning in type-II superconductors~\cite{Blatter1994} or domain-wall pinning by extrinsic disorder~\cite{Lemerle1998,Metaxas2007}, the pinning landscape and stabilized phase defect are \emph{co-generated} by the same microscopic interactions.

Looking forward, this stabilization mechanism could enable dark solitons to serve as phase-sensitive interferometric elements and to underpin weak-link and phase-slip dynamics in atomtronic circuits in higher dimensions, extending their utility beyond 1D geometries~\cite{Scott2008,Haug2018,Polo2019}. In addition, the response of a moving dark soliton to a periodic density background could provide an experimentally feasible protocol for  assessing the rigidity of the SS, directly complementing existing probes~\cite{Mukherjee2023LinChain, SenarathShear2025, Blakie2025, Bougas2026}. Experimentally, this stabilization mechanism could be tested in dipolar SS and spin-orbit-coupled SS states~\cite{Li2017,Leonard2017, Chisholm2026Science}, and may extend to other platforms, such as dipolar molecules~\cite{Zhang2026DipolarMoleculeDroplets}, and exciton-polariton condensates~\cite{Trypogeorgos2025}, provided that stripe order is present.

 {\it Acknowledgments}---This work was supported by JSPS KAKENHI Grant Nos. JP23K03276, JP25KF0135, and JP26K00638. K.M. acknowledges financial support through the JSPS Postdoctoral Fellowship (Fellowship No. P25029). K. M. thanks Malte Schubert for a careful reading of the manuscript and for his helpful comments.

\bibliography{reference}

\section{End Matter}
\prlsection{Appendix A: Particle-Number Scaling and Soliton Stabilization by Stripe-SS Order}

\begin{figure}[t]
    \centering
    \includegraphics[width=\columnwidth]{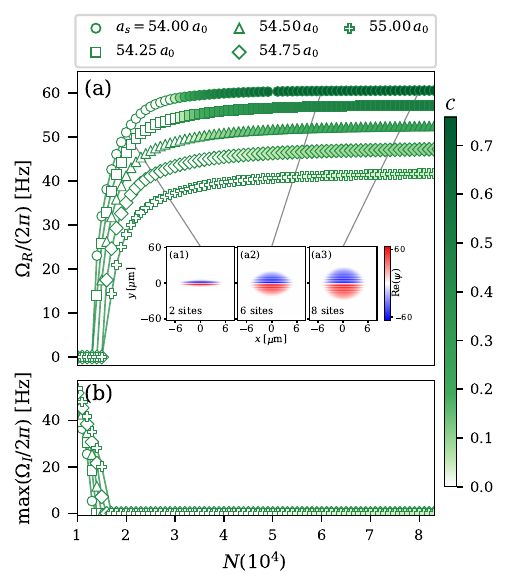}
\caption{
(a) Real part, $\Omega_R$, and (b) maximum imaginary part, $\text{max}[\Omega_I]$, of the excitation frequency as functions of particle number $N$ for several scattering lengths $a_s$ (see legends). The Thomas-Fermi density, $\rho \sim N\omega_x\omega_y \sim 60000\omega_y$, is kept constant by varying the trapping frequency $\omega_x$ simultaneously. Each data point is color-coded according to the value of the density contrast. Insets (a1), (a2), and (a3) show representative solitonic stationary states with 2, 6, and 8 localized density peaks, respectively, at $a_s=54a_0$ for the particle numbers indicated by the black lines.
} \label{fig:appendix1}

\end{figure}

In the main text, we demonstrated the enhanced stability of a straight soliton embedded in a 2D SS  phase, focusing on a benchmark case with a fixed atom number $N = 40000$ and a trapping geometry that yields a four-site localized density modulation. To assess the generality of this stabilization mechanism and confirm its broad experimental feasibility across a wider parameter space, we systematically analyze the collective excitation spectrum under variations of the total particle number, considering $\alpha = \pi/2$. 

Specifically, we preserve the characteristic Thomas-Fermi density scale, $\rho\sim N\omega_x\omega_y$, by varying $N$ while imposing $N\omega_x=40000\times7.5\,\mathrm{Hz}$ and adjusting the trapping frequencies to maintain the same underlying density $\rho$. This allows us to examine the role of the self-induced periodic landscape for various density sites on the transverse excitation spectrum, as illustrated in Fig.~\ref{fig:appendix1}. This scaling procedure effectively controls the number of localized density sites within the SS state, following an approach previously used to characterize dipolar SS in the absence of a soliton~\cite{poli2021}. Indeed, we systematically realize states containing 2, 6, and 8 localized density peaks, shown exemplarily in the insets of Fig.~\ref{fig:appendix1}(a1)--(a3), for total atom numbers $N=20000$, $60000$, and $80000$, respectively.

At low atom numbers, the system remains in the unmodulated SF regime, with vanishing density contrast ($\mathcal{C}=0$). In this regime, the transverse soliton mode exhibits a sizable imaginary component $\Omega_I$ [Fig.~\ref{fig:appendix1}], signaling a strong dynamical transverse instability. As $N$ increases, $\Omega_I$ is rapidly suppressed and vanishes at a critical threshold $N_c$, beyond which the collective spectrum becomes purely real. Simultaneously, the emergence of crystalline stripe order, quantified by the growing density contrast $\mathcal{C}$, hardens the transverse mode and stabilizes the straight soliton.

Therefore, this scaling analysis confirms that the pronounced SS density modulation perpendicular to the soliton stabilizes the defect by enhancing the effective stiffness, as discussed in the main text. This highlights a self-organized mechanism for suppressing the snake instability and stabilizing planar topological defects in dipolar SS.

\begin{figure}[t]
\centering
\includegraphics[width=\columnwidth]{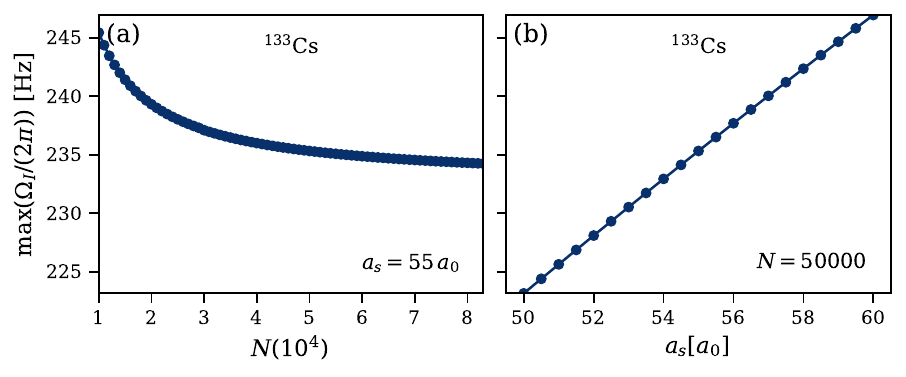}
\caption{
Maximum imaginary BdG frequency, $\max[\Omega_I]/(2\pi)$, for a straight soliton embedded in a $^{133}$Cs condensate: (a) as a function of particle number $N$ at fixed $a_s=55\,a_0$, and (b) as a function of scattering length $a_s$ at fixed $N=50000$.
}
\label{fig:appendix2}
\end{figure}

\prlsection{Appendix B: Soliton Instability in Purely Contact-Interacting Condensates}
The stabilization mechanism elucidated in the main text crucially relies on the long-range dipolar interaction, which amplifies the local density gradients surrounding the soliton core and establishes a robust pinning landscape. To definitively isolate the necessity of these nonlocal interactions and demonstrate that short-range contact interactions alone are insufficient to sustain a stable planar defect, we examine a benchmark system consisting of a purely non-dipolar $^{133}\mathrm{Cs}$ condensate~\cite{Weber2003}. 

To ensure a rigorous comparison, we maintain the identical mass-density scale, $M\rho$, as employed for the $^{166}\mathrm{Er}$ condensate investigated in the main text. We systematically map the  collective excitation spectra under two distinct configurations: first, as a function of the total particle number $N$ at a fixed scattering length $a_s = 55\,a_0$ [following the scaling methodology of Appendix A], and second, as a function of $a_s$, which is also tunable~\cite{Chin2004},  at a fixed atom number $N = 50000$. Crucially, both the nonlocal dipolar mean field ($\Phi_{\rm dd} = 0$) and the beyond-mean-field LHY correction ($g_{\rm LHY} = 0$) are omitted in these simulations.

Throughout the entire equivalent parameter space scanned, the real part of the lowest transverse collective excitation frequency remains  zero. Figure~\ref{fig:appendix2} displays the corresponding maximum imaginary frequency component, $\max[\Omega_I]$. Notably, the  $\Omega_I$ is significantly larger than that observed in the dipolar system, demonstrating that the straight soliton is highly unstable across all considered parameter space. This comparison shows that short-range contact interactions alone do not stabilize a straight soliton in the present 2D geometry. Stabilization instead relies on the nonlocal dipolar interaction, whose effect is enhanced by the self-organized crystalline landscape of the SS phase, as shown in the main text.

 \newpage

\clearpage
\onecolumngrid

\begin{center}

{\large\bfseries Supplemental Material: Self-Organized Stabilization of Straight Dark Solitons 
in a Stripe Supersolid}\\[1em]

Koushik Mukherjee and Hiroki Saito\\[0.5em]

Department of Engineering Science, University of Electro-Communications, 
Tokyo 182-8585, Japan
\end{center}

\vspace{1em}

\vspace{1em}

\twocolumngrid

\renewcommand{\thesection}{S\arabic{section}}
\renewcommand{\thesubsection}{S\arabic{section}.\arabic{subsection}}
\setcounter{section}{0}

\section{S1.~Specifications of the Technical Framework}

\subsection{Effective Quasi-2D Extended Gross--Pitaevskii Equation}

The dynamics of the in-plane condensate wave function $\psi(\vb{r},t)$ with $\vb{r}=(x,y)$, normalized according to $\int d\vb{r}\,|\psi(\vb{r},t)|^2 = 1$, are determined by the effective quasi-two-dimensional (quasi-2D) extended Gross--Pitaevskii equation (eGPE)~\cite{Fischer2006, Ripley2026}
\begin{equation}\label{eq_eGPE_supp}
\begin{aligned}
i\hbar\frac{\partial \psi(\vb{r},t)}{\partial t}
=
\bigg[
&-\frac{\hbar^2\nabla_{\vb{r}}^2}{2M}
+ V(\vb{r})
+ g|\psi(\vb{r},t)|^2  \\
&+ g_{\mathrm{dd}}\Phi_{\mathrm{dd}}(\vb{r},t)
+ g_{\mathrm{LHY}}|\psi(\vb{r},t)|^3
\bigg]\psi(\vb{r},t).
\end{aligned}
\end{equation}
The spatial trapping potential is defined as $V(\vb{r})=\frac{1}{2}M\omega_y^2(\kappa^2 x^2 + y^2)$, where $\kappa=\omega_x/\omega_y$ represents the in-plane trap anisotropy. The effective contact and nominal dipolar interaction strengths are given by $g=\hbar^2\sqrt{8\pi\lambda}\,N a_s / (l_yM)$ and $g_{\mathrm{dd}}=\hbar^2\sqrt{8\pi\lambda}\,N a_{\mathrm{dd}}/(l_yM)$, respectively, expressed in terms of the axial oscillator length $l_y=\sqrt{\hbar/(M\omega_y)}$ and the axial-to-transverse confinement ratio $\lambda=\omega_z/\omega_y$. The characteristic dipolar length is defined as $a_{\mathrm{dd}} = \mu_0 \mu_{\rm m}^2 M / (12\pi\hbar^2)$, where $\mu_{\rm m}$ is the magnetic dipole moment. The beyond-mean-field quantum fluctuation contribution is parameterized by the Lee--Huang--Yang (LHY) coupling constant
\begin{equation}
g_{\mathrm{LHY}}=\frac{\hbar^2128\sqrt{\pi}}{3M}\sqrt{\frac{2}{5}}\left(\frac{\lambda}{\pi}\right)^{3/4}N^{3/2}\left(\frac{a_s}{l_y}\right)^{5/2}\left(1+\frac{3}{2}\varepsilon_{\mathrm{dd}}^2\right),
\end{equation}
where $\varepsilon_{\mathrm{dd}}=a_{\mathrm{dd}}/a_s$ denotes the relative interaction ratio. The total energy functional $E[\psi]$ corresponding to Eq.~\eqref{eq_eGPE_supp} reads
\begin{equation}\label{eq_eGPE_energy}
\begin{aligned}
E[\psi] =
&\int d\mathbf{r}\,
\left[
\frac{\hbar^2}{2M}|\nabla\psi|^2
+V(\mathbf{r})|\psi|^2
+\frac{g}{2}|\psi|^4
+\frac{2}{5}g_{\rm LHY}|\psi|^5
\right] \\
&+\frac{g_{\rm dd}}{2}
\iint d\mathbf{r}\,d\mathbf{r}'\,
U_{\rm dd}^{2{\rm D}}(\mathbf{r}-\mathbf{r}')
|\psi(\mathbf{r})|^2|\psi(\mathbf{r}')|^2 .
\end{aligned}
\end{equation}

\subsection{Quasi-2D Dipolar Kernel and Momentum-Space Evaluation}

The nonlocal dipolar potential is evaluated via spatial convolution, $\Phi_{\mathrm{dd}}(\vb{r},t) = \int d\vb{r}'\, U_{\mathrm{dd}}^{2\mathrm{D}}(\vb{r}-\vb{r}') |\psi(\vb{r}',t)|^2$, which is computed efficiently in momentum space as $\tilde{\Phi}_{\mathrm{dd}}(\mathbf{k},t)=\tilde{U}_{\mathrm{dd}}^{2\mathrm{D}}(\mathbf{k})\,\tilde{n}(\mathbf{k},t)$, where $\tilde{n}(\mathbf{k},t)=\mathcal{F}[|\psi(\vb{r},t)|^2]$ is the Fourier transform of the 2D particle density. For dipoles polarized within the $x$--$z$ plane at an angle $\alpha$ relative to the $z$ axis, the effective quasi-2D interaction kernel is given by~\cite{Fischer2006, Ronen2006}
\begin{equation}
\tilde{U}_{\mathrm{dd}}^{2\mathrm{D}}(\mathbf{k}) = \cos^2\alpha\,h_{\perp}(\mathbf{q}) + \sin^2\alpha\,h_{\parallel}(\mathbf{q}),
\end{equation}
where the anisotropic geometric scaling functions are defined using the dimensionless momentum components $q_x=k_xl_{z}/\sqrt{2}$, $q_y=k_yl_{z}/\sqrt{2}$ [with $l_z = \sqrt{\hbar/(M\omega_z)}$] and $q=\sqrt{q_x^2+q_y^2}$ as
\begin{align}
h_{\perp}(\mathbf{q}) &= 2-3\sqrt{\pi}\,q\,e^{q^2}\mathrm{erfc}(q), \\
h_{\parallel}(\mathbf{q}) &= -1+3\sqrt{\pi}\,\frac{q_x^2}{q}e^{q^2}\mathrm{erfc}(q).
\end{align}

\subsection{Numerical Implementation}

The stationary dark-soliton states $\psi_0(\vb{r})$ are found by evolving Eq.~\eqref{eq_eGPE_supp} in imaginary time ($\tau = -it$). To structurally pin the planar defect along the nodal line $y=0$, the phase constraint $\psi(x,y,\tau) \to \mathrm{sgn}(-y)\abs{\psi(x,y,\tau)}$ is strictly enforced after each discrete imaginary time step $\Delta \tau$. Real-time dynamics is simulated by direct forward-time propagation of Eq.~\eqref{eq_eGPE_supp}. Both numerical regimes utilize a second-order split-step Fourier spectral method within a computational box of size $L_x = L_y = 100 \mu m$. We confirmed numerical convergence by comparing results across grid resolutions of $512 \times 512$, and $512 \times 256$ grid points.

\subsection{Linearized Bogoliubov--de Gennes Operators}

Linearization of Eq.~\eqref{eq_eGPE_supp} around the stationary background profile $\psi_0(\vb{r})$ via the standard Bogoliubov ansatz $\psi(\vb{r},t)=e^{-i\mu t/\hbar} [ \psi_0(\vb{r}) + \epsilon ( u(\vb{r})e^{-i\Omega t} + v^*(\vb{r})e^{i\Omega t} ) ]$ leads to the coupled matrix eigenvalue problem
\begin{equation}\label{eq_BdG_matrix_supp}
\begin{bmatrix}
L_{11} & L_{12} \\
L_{21} & L_{22}
\end{bmatrix}
\begin{bmatrix}
u(\vb{r}) \\
v(\vb{r})
\end{bmatrix}
= \hbar\Omega
\begin{bmatrix}
u(\vb{r}) \\
v(\vb{r} )
\end{bmatrix}.
\end{equation}
The linear matrix operators satisfy the symmetry constraints $L_{22}=-L_{11}$ and $L_{21}=-L_{12}^*$. The explicit real-space forms of the coupled differential blocks acting on a general function $f(\mathbf{r})$ are given by
\begin{align}
L_{11}f(\mathbf{r}) = 
&\left[-\frac{\hbar^2\nabla_{\mathbf{r}}^2}{2M} 
+ V(\mathbf{r}) - \mu + 2g|\psi_0|^2\right.\nonumber\\
&\left.+ g_{\mathrm{dd}}\Phi_{\mathrm{dd}}^{(0)}(\mathbf{r}) 
+ \frac{5}{2}g_{\mathrm{LHY}}|\psi_0|^3 \right]f(\mathbf{r}) 
\nonumber\\
&+ g_{\mathrm{dd}}\psi_0(\mathbf{r})
\int d\mathbf{r}'\,U_{\mathrm{dd}}^{2\mathrm{D}}(\mathbf{r}-\mathbf{r}')
\psi_0^*(\mathbf{r}')f(\mathbf{r}'), 
\label{eq:L11_explicit} \\[6pt]
L_{12}f(\mathbf{r}) = 
&\left[g\psi_0^2 
+ \frac{3}{2}g_{\mathrm{LHY}}|\psi_0|\psi_0^2 
\right]f(\mathbf{r}) \nonumber\\
&+ g_{\mathrm{dd}}\psi_0(\mathbf{r})
\int d\mathbf{r}'\,U_{\mathrm{dd}}^{2\mathrm{D}}(\mathbf{r}-\mathbf{r}')
\psi_0(\mathbf{r}')f(\mathbf{r}'). 
\label{eq:L12_explicit}
\end{align}
Here, $\Phi_{\mathrm{dd}}^{(0)}(\mathbf{r}) = \int d\mathbf{r}'\, U_{\mathrm{dd}}^{2\mathrm{D}}(\mathbf{r}-\mathbf{r}') |\psi_0(\mathbf{r}')|^2$ denotes the static mean-field dipolar potential evaluated for the unperturbed dark-soliton state. The resulting large-scale sparse BdG matrix in Eq.~\eqref{eq_BdG_matrix_supp} is diagonalized using the iterative eigensolver routines implemented in the \textsc{Spectra} library~\cite{Spectra}.

\section*{S2.\ Derivation of the Bending Energy}
We derive the excess energy associated with a stationary transverse deformation of the soliton, providing the analytical basis for the stiffness coefficient $\sigma_2$ discussed in the main text. During the dynamics, an unstable soliton typically passes through such intermediate bent configurations before forming final nonlinear structures, such as solitonic vortices or vortex--antivortex pairs. Our model explicitly demonstrates how the SS density modulation stiffens the soliton against bending, making both the intermediate bent states and the resulting nonlinear structures more difficult to access dynamically, and thus enhancing the stability of stationary straight solitons. In this way, it complements the spectral analysis used to assess soliton stability.

We introduce a weak transverse displacement $u(x)=A\sin(kx)$, with $x\in[-L_x/2,L_x/2]$, and approximate the deformed density and phase profiles by local shifts of the stationary solution, $n(x,y)=n_0\left(y-u(x)\right)$ and
$\theta(x,y)=\theta_0\left(y-u(x)\right)$. Expanding to second order in $u$ gives
\begin{equation}
   \delta n(x,y)
    = -u(x)\,\partial_y n_0
    + \frac{u^2(x)}{2}\,\partial_y^2 n_0,
\label{eq:SM_delta_n}
\end{equation}
and
\begin{equation}
    \delta\theta(x,y)
    = -u(x)\,\partial_y\theta_0
    + \frac{u^2(x)}{2}\,\partial_y^2\theta_0
\label{eq:SM_delta_theta}
\end{equation}

Two $x$-integrals of the sinusoidal modulation appear repeatedly below:
\begin{equation}
\int_{-L_x/2}^{L_x/2}\!dx\;u^2
=\frac{A^2L_x}{2}\!\left[1-\frac{\sin (kL_x)}{kL_x}\right],
\label{eq:SM_I1x}
\end{equation}
\begin{equation}
\int_{-L_x/2}^{L_x/2}\!dx\;(\partial_x u)^2
=\frac{A^2L_xk^2}{2}\!\left[1+\frac{\sin (kL_x)}{kL_x}\right].
\label{eq:SM_I2x}
\end{equation}

\textit{\textbf{Kinetic energy}.}
We first derive the kinetic-energy cost of soliton bending. The kinetic part of the eGPE energy functional, Eq.~\ref{eq_eGPE_energy},
 can be decomposed into amplitude and phase contributions as
\begin{equation}
E_{\mathrm{kin}} = \frac{\hbar^2}{2M} \int d\mathbf{r} \left[~|\nabla \sqrt{n_{0}}|^2 + n_{0} |\nabla \theta|^2 ~\right].
\end{equation}

We first focus on the quantum-pressure contribution to the kinetic energy:
\begin{equation}\nonumber
E_{\mathrm{kin}}^{(q)} = \frac{\hbar^2}{2M} \int d\mathbf{r} \, |\nabla \sqrt{n_{0}}|^2.
\end{equation}
The corresponding change in the kinetic energy due to a bent soliton is
\begin{equation}
\begin{aligned}
\Delta E_{\mathrm{kin}}^{(q)}
&= E_{\mathrm{kin}}^{(q)}(n_{0}+\delta n)
   -E_{\mathrm{kin}}^{(q)}(n_{0}) \\[6pt]
&= \frac{\hbar^2}{2M}\int d\mathbf{r}\,
\left|
\nabla \frac{\delta n(x,y)}{2\sqrt{n_{0}}}
\right|^2 .
\end{aligned}
\end{equation}

Using Eq.~\ref{eq:SM_delta_n} and $f = \sqrt{n_{0}}$, the above equation can be cast into the form,
\begin{align}
\Delta E_{\mathrm{kin}}^{(q)} 
&= \frac{\hbar^2}{8M} \Bigg\{
\left[\int dx\, (\partial_x u)^2\right]
\left[\int dy\, (\partial_y f)^2 \right] \nonumber \\
&\quad -
\left[\int dx\, u(x)^2\right]
\left[\int dy\, 
\frac{(\partial_y f)^2\,\partial_y^2 f}{f}
\right]
\Bigg\}.
\end{align}.

Substituting the explicit integrals, Eq.~\ref{eq:SM_I1x} and ~\ref{eq:SM_I2x}, and further assuming $f = \chi^2$ to simplify the second term in the above equation, we get the final quantum-pressure contribution to the bending energy cost:
\begin{equation}\label{eq:kinEn_QP}
\begin{aligned}
\Delta E_{\mathrm{kin}}^{(q)}
=
&\frac{\hbar^2 A^2 L_x}{4M}
\Bigg\{
 k^2
\left[
1 + \frac{\sin(k L_x)}{kL_x}
\right]
\int dy\,
\left(
\partial_y \sqrt{n_0}
\right)^2
\\[6pt]
& -8
\left[
1 - \frac{\sin(k L_x)}{kL_x}
\right]
\int dy\,
\bigg [
\left(\partial_y n_0^{1/4}\right)^4 + \\
&
n_0^{1/4}
\left(\partial_y n_0^{1/4}\right)^2
\partial_y^2 n_0^{1/4}
\bigg ]
\Bigg\}.
\end{aligned}
\end{equation}

Next we focus on the contribution of the phase-gradient term to the bending energy cost,
\begin{align}\label{eq:kinetic_SM}
\Delta E^{(p)}_{\mathrm{kin}}
&=
\frac{\hbar^2}{2M}\int d\mathbf{r}\,\bigg[
\bigl(n_{0}+\delta n\bigr)
\left|\nabla(\theta+\delta\theta)\right|^2
-
n_{0}\left|\nabla\theta\right|^2 \bigg],
\end{align}

which, utilizing the Eqs.~\ref{eq:SM_delta_n} and \ref{eq:SM_delta_theta}, and performing some algebra, reduces to 
\begin{equation}
\begin{aligned}
\Delta E^{(p)}_{\mathrm{kin}}
=
 \frac{\hbar^2 A^2 L_x k^2}{4M}
    \left[1 + \frac{\sin(kL_x)}{k L_x}\right]
    \int dy\;n_0(y)\,(\partial_y\theta_0)^2.
\end{aligned}
\end{equation}

For a stationary dark soliton, the continuity equation gives $\partial_y \left(n_0 \partial_y \theta_0\right) = 0.$
Therefore, $n_0 \partial_y \theta_0 = j_0,$
where $j_0$ is the background current, which is proportional to the soliton velocity $v_s$.
Therefore, one can get
\begin{equation}
    \partial_y \theta_0 = \frac{j_0}{n_0} 
    \qquad
    \partial^2_y \theta_0 = -\frac{j_0 \partial_y n_0}{n^2_0}
\end{equation}

For a dark soliton, $v_s = 0$, and hence $j_0 = 0$. Therefore,
\begin{equation}
    \Delta E^{(p)}_{\mathrm{kin}} = 0 .
\end{equation}

Thus, the phase-gradient contribution vanishes; the remaining bending cost comes from the quantum-pressure, contact, LHY, and dipolar terms.\\

\textit{\textbf{Potential Energy.}}
In our  model, we neglect the density variation along the $x$-axis, and thus the bending energy cost is contributed by the trap potential $V(y) = M\omega^2 y^2/2$
\begin{equation}
        \Delta E_{\rm pot} = E_{\rm pot} (n_{0} + \delta n ) - E_{\rm pot} (n_{0})
    = \int d\vb{r}\;V(y)\,\delta n(x,y).
\end{equation}
Substituting Eq.~\eqref{eq:SM_delta_n}, we split this into two contributions:
\begin{equation*}
\begin{aligned}
\Delta E_{\rm pot}
&=
-\int dx\, u(x) \int dy\, V(y)\,\partial_y n_0(y) \\
&\quad
+ \int dx\,u^2(x)\; \int dy\, \frac{V(y)}{2}\,\partial_y^2 n_0(y).
\end{aligned}
\end{equation*}

The first term in the above equation vanishes, and then utilizing the Eq.~\eqref{eq:SM_I1x}, we arrive at

\begin{equation}
   \Delta E_{\rm pot} =  \frac{A^2L_x}{4}\!\left[1-\frac{\sin (kL_x)}{kL_x}\right]\;\int dy\, V(y)\,\partial_y^2 n_0(y).
\end{equation}
\\
\textit{\textbf{Contact interaction.}}
From $E_{\rm c}=(g/2)\int d\mathbf{r}\,n_{0}^2$, in Eq.~\ref{eq_eGPE_energy}, one can calculate the bending-energy cost due to contact interaction,
\begin{equation}
    \Delta E_c = E_c(n_{0} + \delta n) - E_c(n_{0}) = g\int d\mathbf{r}\;n_{0}\delta n + \frac{g}{2} \int d\mathbf{r}\; \delta n^2.
\end{equation}

In the first term, $\mathcal{O}(u)$ part integrates to zero. 
Considering the $\mathcal{O}(u^2)$ part,  we integrate by parts and get
\begin{equation}\label{eq:c_first}
    \frac{g}{2}\int dx\,u^2\int dy\; n_0\,\partial_y^2 n_0.
=
    -\frac{g}{2}\int dx\,u^2\int dy\;(\partial_y n_0)^2.
\end{equation}

In the second term, expanding $(\delta n)^2$ and retaining only $\mathcal{O}(u^2)$, we get
\begin{equation}\nonumber
    (\delta n)^2
    = \left[-u\,\partial_y n_0 + \frac{u^2}{2}\,\partial_y^2 n_0\right]^2
    = u^2(\partial_y n_0)^2 + \mathcal{O}(u^3).
\end{equation}
Therefore,
\begin{equation}\label{eq:c_second}
    \frac{g}{2}\int dx\,dy\;(\delta n)^2
    = \frac{g}{2}\int dx\,u^2\int dy\;(\partial_y n_0)^2.
\end{equation}
Adding \eqref{eq:c_first} and \eqref{eq:c_second}:

\begin{equation}
\Delta E_{\mathrm{c}}
=
0.
\label{eq:SM_Ec}
\end{equation}
\\
\textit{\textbf{LHY correction.}}
The LHY contribution to the bending energy cost can be written as
\begin{equation}\nonumber
    \Delta E_{\rm LHY} = \frac{2}{5}g_{\rm LHY} \int d\mathbf{r}\, \big( (n_{0} + \delta n)^{5/2} - n^{5/2}_{0}\big)
\end{equation}
Expanding $(n_0+\delta n)^{5/2}$ to second order in $\delta n$,
\begin{equation}\nonumber
(n_0+\delta n)^{5/2}
=n_0^{5/2}
+\frac{5}{2}n_0^{3/2}\,\delta n
+\frac{15}{8}n_0^{1/2}\,(\delta n)^2
+\mathcal{O}(\delta n^3),
\end{equation}
we arrive at 
\begin{equation}\label{eq:lhy_cost}
    \Delta E_{\mathrm{LHY}}
    = g_{\mathrm{LHY}}\int dx\,dy\; n_0^{3/2}\,\delta n
    + \frac{3}{4}g_{\mathrm{LHY}}\int dx\,dy\; n_0^{1/2}(\delta n)^2.
\end{equation}
In Eq.~\ref{eq:lhy_cost}, the $\mathcal{O}(u)$ part in the first term again vanishes upon $x$-integration, and the $\mathcal{O}(u^2)$ contribution from the first term is
\begin{equation}\label{eq:lhy_first}
    -\frac{3g_{\mathrm{LHY}}}{4}\int dx\,u^2
    \int dy\; n_0^{1/2}(\partial_y n_0)^2.
\end{equation}

Retaining only $\mathcal{O}(u^2)$ in $(\delta n)^2$ in the second term of Eq.~\ref{eq:lhy_cost}, we get
\begin{equation}
    (\delta n)^2 = u^2(\partial_y n_0)^2 + \mathcal{O}(u^3).
\end{equation}
Therefore,
\begin{equation}\label{eq:lhy_second}
    \frac{3g_{\mathrm{LHY}}}{4}\int dx\,dy\; n_0^{1/2}(\delta n)^2
    = \frac{3g_{\mathrm{LHY}}}{4}\int dx\,u^2
    \int dy\; n_0^{1/2}(\partial_y n_0)^2.
\end{equation}

Adding \eqref{eq:lhy_first} and \eqref{eq:lhy_second}:

\begin{equation}\label{eq:EnLhy_SM}
\begin{aligned}
\Delta E_{\mathrm{LHY}}
= 0.
\end{aligned}
\end{equation}

\textit{\textbf{Dipolar interaction.}}
In the momentum space, the dipolar interaction energy, from Eq.~\ref{eq_eGPE_energy}, can be written as, using Parseval's theorem,
\begin{equation}
    E_{\rm dd}
    =
    \frac{g_{\rm dd}}{2}
    \int\frac{d\mathbf{k}}{(2\pi)^2}\,
    \widetilde{U}_{\rm dd}^{\rm 2D}(\mathbf{k})\,
    |\widetilde{n}(\mathbf{k})|^2.
\end{equation}
The energy cost of bending is obtained by subtracting the straight-soliton
contribution:
\begin{equation}
    \Delta E_{\rm dd}
    =
    \frac{g_{\rm dd}}{2}
    \int\frac{d\mathbf{k}}{(2\pi)^2}\,
    \widetilde{U}_{\rm dd}^{\rm 2D}(\mathbf{k})
    \left[
    |\widetilde{n}_0(\mathbf{k}) + \widetilde{\delta n}(\mathbf{k})|^2
    -
    |\widetilde{n}_0(\mathbf{k})|^2
    \right].
\end{equation}
Expanding the modulus squared:
\begin{equation}
    |\widetilde{n}_0 + \widetilde{\delta n}|^2 - |\widetilde{n}_0|^2
    =
    2\,\mathrm{Re}\!\left[\widetilde{n}_0^*\,\widetilde{\delta n}\right]
    +
    |\widetilde{\delta n}|^2,
\end{equation}
so we write
\begin{equation}\label{eq:dipolat_SM}
    \Delta E_{\rm dd}
    =
    \Delta E_{\rm dd}^{(1)}
    +
    \Delta E_{\rm dd}^{(2)},
\end{equation}
where
\begin{align}
    \Delta E_{\rm dd}^{(1)}
    &=
    g_{\rm dd}
    \int\frac{d\mathbf{k}}{(2\pi)^2}\,
    \widetilde{U}_{\rm dd}^{\rm 2D}(\mathbf{k})\,
    \mathrm{Re}\!\left[\widetilde{n}_0^*(k_y)\,\widetilde{\delta n}(k_x,k_y)\right],
    \\[6pt]
    \Delta E_{\rm dd}^{(2)}
    &=
    \frac{g_{\rm dd}}{2}
    \int\frac{d\mathbf{k}}{(2\pi)^2}\,
    \widetilde{U}_{\rm dd}^{\rm 2D}(\mathbf{k})\,
    |\widetilde{\delta n}(k_x,k_y)|^2.
\end{align}

In the above expressions, we have,
\begin{equation}\label{eq:delta_n_FT}
    \widetilde{\delta n}(k_x,k_y)
    =
   -ik_y\,\widetilde{u}(k_x)\,\widetilde{n}_0(k_y)
    -\frac{k_y^2}{2}\,\widetilde{u^2}(k_x)\,\widetilde{n}_0(k_y),
\end{equation}
where $\widetilde{u}(k_x)$ is the Fourier transform of $u(x)$ and
$\widetilde{u^2}(k_x)$ is the Fourier transform of $u^2(x)$.

In particular, 
\begin{equation}
\widetilde{u}(k_x)
=
\frac{A L_x}{2i}
\left\{
\frac{\sin\!\left[(k_x-k)L_x/2\right]}{(k_x-k)L_x/2}
-
\frac{\sin\!\left[(k_x+k)L_x/2\right]}{(k_x+k)L_x/2}
\right\},
\end{equation}
and by convolution, 
\begin{equation}
    \widetilde{u^2}(k_x)
    =
    \int\frac{dq_x}{2\pi}\,\widetilde{u}(q_x)\,\widetilde{u}(k_x-q_x).
\end{equation}

 Combining the nonvanishing quantum-pressure, trap, and dipolar contributions yields Eq.~\eqref{eq:Gamma} in the main text.

\section*{S3.\ Structure-factor and the growth rate calculation}

We independently determine the instability growth rate from real-time
eGPE simulations by tracking the exponential amplification of the density
modulation generated by the unstable transverse mode. For each scattering
length, the perturbed time evolution is initialized from the stationary
straight-soliton state. The density fluctuation at time $t$ reads
\begin{equation}
\Delta n(\mathbf{r},t)=n(\mathbf{r},t)-n(\mathbf{r},0).
\end{equation}
Then the structure factor is calculated as
\begin{equation}
S(k_x,k_y,t)=
\left|
\mathcal{F}_{2D}\!\left[
\Delta n(x,y,t)
\right](k_x,k_y)
\right|^2 .
\label{eq:SM_structure_factor}
\end{equation}
At each time we select the maximum nonzero Fourier-power value,
$S_{\max}(t)=\max_{(k_x,k_y)\ne(0,0)}S(k_x,k_y,t)$. We then fit
\begin{equation}
S_{\max}(t)=S_0\exp(2\gamma_{\rm rt}t),
\label{eq:SM_growth_fit}
\end{equation}
over the early-time interval where $\log S_{\max}(t)$ is linear. The factor
of two appears because $S$ is a density-power spectrum, whereas
$\gamma_{\rm rt}$ is the growth rate of the perturbation amplitude. For
the ensemble data (51 independent simulations in our case), the same procedure is applied independently to each
noise realization and the reported value is the arithmetic mean of the
resulting $\gamma_{\rm rt}/(2\pi)$, which is  used in Fig.~\ref{fig:fig2_manuscript}(c).

\end{document}